# LOCld65, A Dual-Channel VCSEL Driver ASIC For Detector Front-End Readout

Wei Zhou, Datao Gong, Quan Sun, Di Guo, Guangming Huang, Kai Jin, Chonghan Liu, Jun Liu, Tiankuan Liu, Ming Qi, Xiangming Sun, James Thomas, Le Xiao, Jingbo Ye

*Abstract*—We present the design and the test results of a dual-channel Vertical-Cavity Surface-Emitting Laser (VCSEL) driver ASIC LOCld65 for detector front-end readout. LOCld65 is designed in a commercial 65-nm CMOS technology with a power supply of 1.2 V. LOCld65 contains two separate channels with the same structure and the two channels share an I²C slave. Each channel consists of an input amplifier, four stages of limiting amplifiers (LAs), a high-current output driver, and a bias-current generator. In order to extend the bandwidth, the input amplifier uses an inductive peaking technique and the LAs use a shared inductive peaking technique. The input amplifier and the output driver each utilize a Continuous-Time Linear Equalizer (CTLE). The LAs employ active feedback. The modulation current, the bias current, the peaking strength of the CTLEs, and the feedback strength of LAs are programmable through an I²C interface. In order to protect from the radiation damage, the I²C slave is implemented with triple modular redundancy. Each channel of LOCld65 is tested to operate up to 14 Gbps with typical power dissipations (the VCSEL included) of 68.3 mW/channel and 62.1 mW/channel at the VCSEL voltages of 3.3 V and 2.5 V, respectively. LOCld65 survives 4.9 kGy(SiO$_2$). LOCld65 is an excellent match for the serializer-deserializer ASIC lpGBT in single- or dual-channel optical transmitters in HL-LHC upgrade applications.

## I. Introduction

WITH the development of detector technologies, the demands for data transmission in high-energy physics experiments have constantly increased. However, the traditional electrical interconnection technology has become an important factor restricting the performance of a large-capacity data transmission system. Benefiting from the advantages of high bandwidth, high channel density, low mass, and no ground loop, a high-speed data transmission system with optical links has been extensively used in readout systems of high-energy physics experiments. A Toroidal LHC Apparatus (ATLAS) and Compact Muon Solenoid (CMS) on the Large Hadron Collider (LHC) use optical links between the front end and the back end [1-2]. The front-end detectors and electronics detect and process information from high-energy particle collisions and transmit the collected data in a harsh radiation environment. In order to ensure proper operation, all components at the front end must meet the radiation tolerance standard proposed by ATLAS [3-5]. Therefore, radiation-tolerant Application-Specific Integrated Circuits (ASICs) are necessary for the front-end electronics of high-energy physics experiments.

We have designed the dual-channel Vertical-Cavity Surface-Emitting Laser (VCSEL) driver, LOCld [6], in a commercial 250-nm Silicon-on-Sapphire CMOS technology. LOCld is the baseline choice for the ATLAS Liquid-Argon-Calorimeter Phase-I upgrade [7-8]. Each channel of LOCld operates at up to 8 Gbps with a power consumption of 200 mW/channel. We upgrade LOCld in data rate and power dissipation by designing an ASIC named LOCld65. LOCld65 is a dual-channel VCSEL driver and each channel operates at a data rate of up to 14 Gbps. LOCld65 is designed in the same 65-nm technology as lpGBT [9], the serializer-deserializer ASIC developed chiefly for the High-Luminosity LHC (HL-LHC) detector upgrades [10]. The serializer of lpGBT operates at 5.12 or 10.48 Gbps. Unlike a VCSEL array driver (such as VLAD [11-12], GBLD10+ [13], LDQ10 [14], and LDQ10+ [15]), which targets at driving a VCSEL array, LOCld65 aims at driving VCSEL Transmitter Optical Sub-Assemblies (TOSAs). Each channel of LOCld65 can be individually turned on or off, making LOCld65 suitable for applications in dual-channel optical transmitters such as MTx [16] and VTTx [17] or in transceivers such as MTRx [16] and VTRx [17]. LOCld65 is an excellent match for lpGBT in single- or dual-channel optical transmitters in HL-LHC upgrade applications.

## II. Design of LOCld65

LOCld65 is designed in a commercial 65-nm CMOS

Manuscript received on June 22, 2018. This work was supported in part by the China Scholarship Council (CSC) and SMU's Dedman College Dean's Research Council Grant. (*Corresponding authors: Tiankuan Liu and Xiangming Sun*)

Wei Zhou is with Central China Normal University, Wuhan, Hubei 430079, China and with Southern Methodist University, Dallas, TX 75205, USA.

Datao Gong, Quan Sun, Chonghan Liu, Tiankuan Liu, James O. Thomas, and Jingbo Ye are with Southern Methodist University, Dallas, TX 75205, USA (Tiankuan Liu: +1-214-768-9204, email: tliu@smu.edu).

Di Guo, Guangming Huang, Kai Jin, Jun Liu, Xiangming Sun, and Le Xiao are with Central China Normal University, Wuhan, Hubei 430079, China (Xiangming Sun: +86-27-67867835, email: xmsun@phy.ccnu.edu.cn).

Ming Qi is with Nanjing University, Nanjing, Jiangsu 210008, China.

technology with a power supply of 1.2 V. LOCld65 has two separate channels with the same structure and a shared I²C slave. Each channel can be individually turned on or off. The design goal is for each channel to amplify a differential signal of greater than or equal to 100 mV (P-P) into an 8-mA modulation current for a VCSEL and to offer a bias current of 6 mA. The minimum input signal amplitude of LOCld65 matches that of commercial VCSEL drivers, such as ONET8501V produced by Texas Instruments, and is less than the output amplitude of a usual serializer, such as lpGBT, to accommodate potential signal loss through traces on Printed Circuit Boards (PCBs) from a serializer to LOCld65.

Each channel consists of an input amplifier, four stages of limiting amplifiers (LAs), a high-current output driver, and a bias-current generator. Fig. 1 shows the block diagram of a channel. The channel has four stages of limiting amplifiers to amplify a voltage signal. The high-current-output driver converts the amplified voltage signal into a modulation current. The output connection of each channel is also shown in the figure. Each channel of LOCld65 is AC-coupled to a VCSEL [18]. Two inductors and two capacitors isolate the bias current and the modulation current. The bias-current generator provides a bias current of from 0 to 8.4 mA.

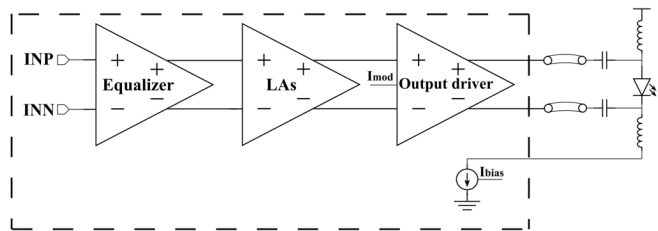

Fig 1. Block diagram of a LOCld65 channel.

*A. Input amplifier*

The input amplifier is a critical function block of the whole channel and should have enough bandwidth to avoid Inter-Symbol Interference. Fig. 2(a) is the schematic of the input amplifier. The input amplifier has two 50-Ω termination resistors in series with the center connected to a common-mode voltage (Vcom shown in the figure). We use two resistors to divide the power voltage of 1.2 V to generate the common-mode voltage, which is 2/3 of the power voltage ($V_{DD}$), independent of temperature change and process variation.

We use a shunt peaking technique to enhance the bandwidth [19]. In the schematic, we use a center-tapped inductor for shunt peaking. Our simulation results show that the bandwidth increases from 8.9 GHz without shunt peaking to 10.8 GHz with shunt peaking.

Considering the high-frequency components of the input signal may be attenuated under exceptional conditions, e.g., with long input traces on a PCB, we design a Continuous-Time Linear Equalizer (CTLE) [20-22] in the input amplifier to compensate the potential high-frequency signal loss. A constant capacitance $C_S$ and a programmable resistor $R_S$ are R-C degeneration of the CTLE. We use N-type MOSFETs connected in parallel to implement the programmable resistor. Fig. 2(b) is the simulated frequency responses of the CTLE with different resistance settings. When the resistance increases, the components at low frequencies are attenuated. As a result, the high-frequency signal loss is compensated.

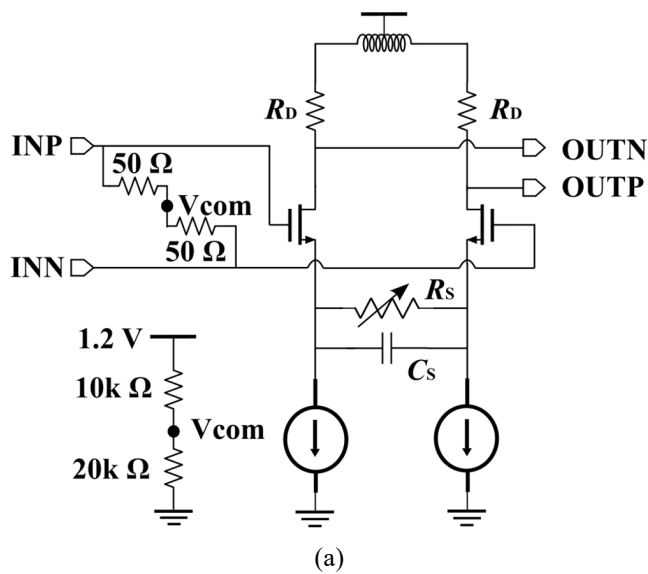

(a)

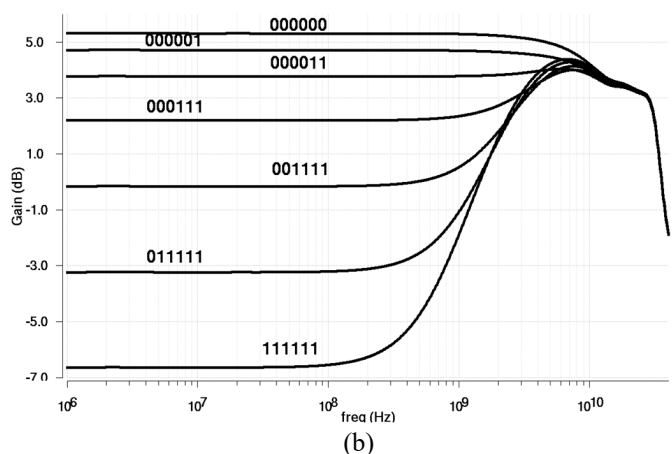

(b)

Fig. 2. Schematics of the input amplifier (a) and simulated amplitude-frequency response curve of the input amplifier (b).

*B. Limiting amplifiers*

The goal of the LA design is to amplify a small voltage signal coming from the input amplifier to an output signal that can fully steer the last-stage tail current of the differential output driver from one arm to the other one to maximize electro-optical conversion efficiency. Therefore, we design the output signal of the LAs with a swing of 800 mV (P-P). Since the input signal swing is as low as 100 mV (P-P), the total gain of the LAs should be no-less-than 18 dB. Since the target data rate is 14 Gbps, the overall bandwidth of the LAs should be greater than 10 GHz, 70% of the data rate (14 Gbps).

Fig. 3(a) is the schematic of the LAs. The number of stages is a trade-off between the gain and the bandwidth. From the first stage to the last stage, the tail current of each stage increases exponentially, whereas the load resistors decrease exponentially. In order to broaden the bandwidth and to reduce chip area, the first and the second stages share an inductor, and the third and the fourth stages share an inductor as well [23, 13].

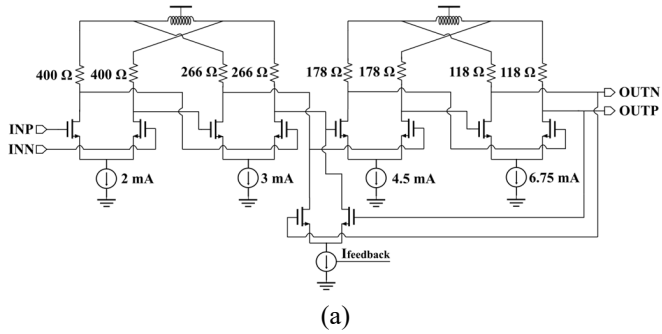

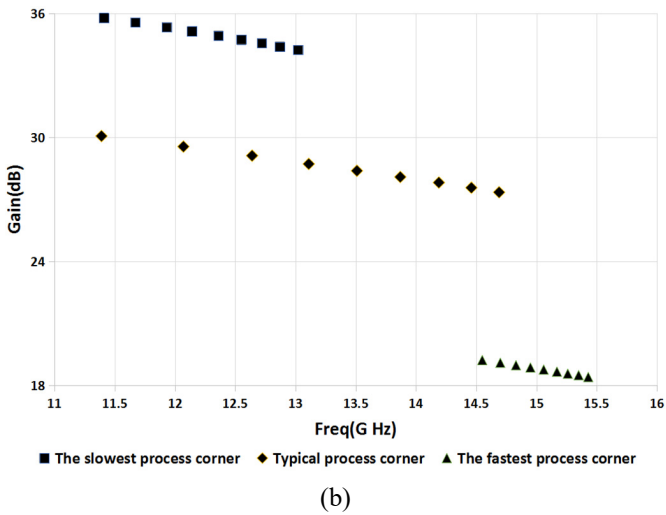

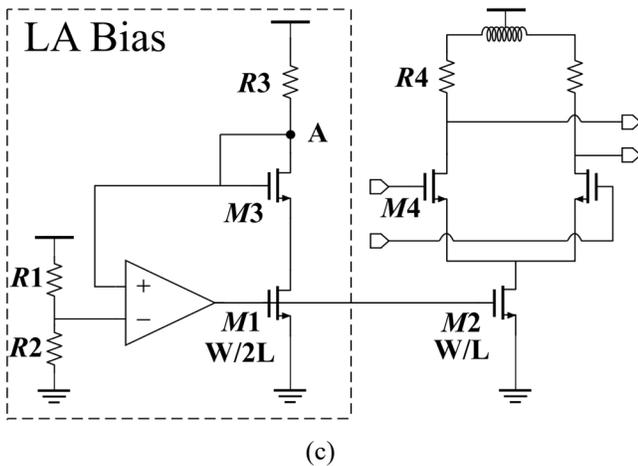

Fig. 3. Schematic of the limiting amplifiers (a), simulated amplitude-frequency response curve of the limiting amplifiers (b), and the bias circuit (c).

It should be noted that the gain and bandwidth of the amplifiers significantly depends on the process variation. The post-layout simulation results show that, at the fastest corner, the LAs meet the requirements of bandwidth and gain, whereas at the slowest corner, the bandwidth is only 5.9 GHz and the gain is higher than 41 dB. To mitigate the process variation, we use an active feedback cell [23-25] between the second stage and the fourth stage. By programming the tail current of the feedback amplifier, the total gain and the overall bandwidth of the limiting amplifiers are greater than 18 dB and 10 GHz, respectively, at all process corners. Fig. 3(b) shows the post-layout simulated gain and bandwidth of the LAs under different feedback current settings and in the different process variations.

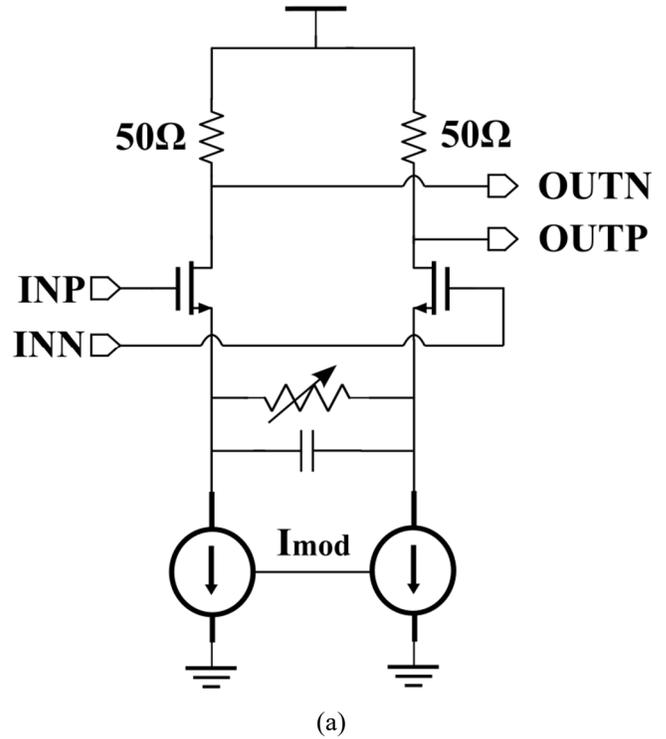

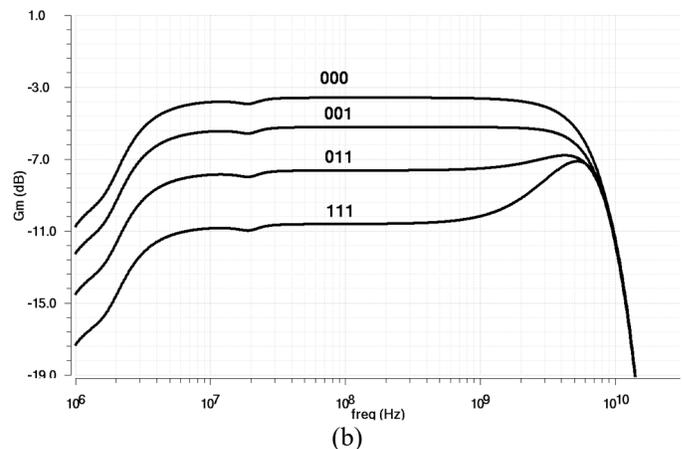

Fig. 4. Schematic of the output driver (a) and post-layout AC simulation results of the output modulation current with different resistance settings (b).

The differential amplifiers are DC coupled, i.e., the input common-mode voltage of each stage is that of its previous stage. Fig. 3(c) shows the first-stage LA and its bias circuit. $R_1$ and $R_2$ are the same types and $R_1$ is a half resistance of $R_2$. The division of $V_{DD}$ with $R_1$ and $R_2$ generates a voltage $\frac{2}{3}V_{DD}$. The op-amp operates as a voltage follower and sets the voltage at Node A ($V_A$) to equal to ⅔$V_{DD}$. $M_1$ has half of the W/L ratio as $M_2$, so the tail current passing through the drain and the source of $M_2$ is twice that of the current passing through the source and the drain of $M_1$. $M_3$ and $M_4$ have the same size. Because of the symmetrical structure, the two arms of the LA uniformly share the tail current $I_{TAIL}$. Therefore, the current passing through $R_4$ is the same as the current passing through $R_3$. The output common-mode voltage is kept at the same voltage of $V_A$, i.e., ⅔$V_{DD}$. By scaling the load resistance and the tail MOSFET size, the common-mode voltage of all LAs is kept at ⅔$V_{DD}$, independent of process variation and temperature change.

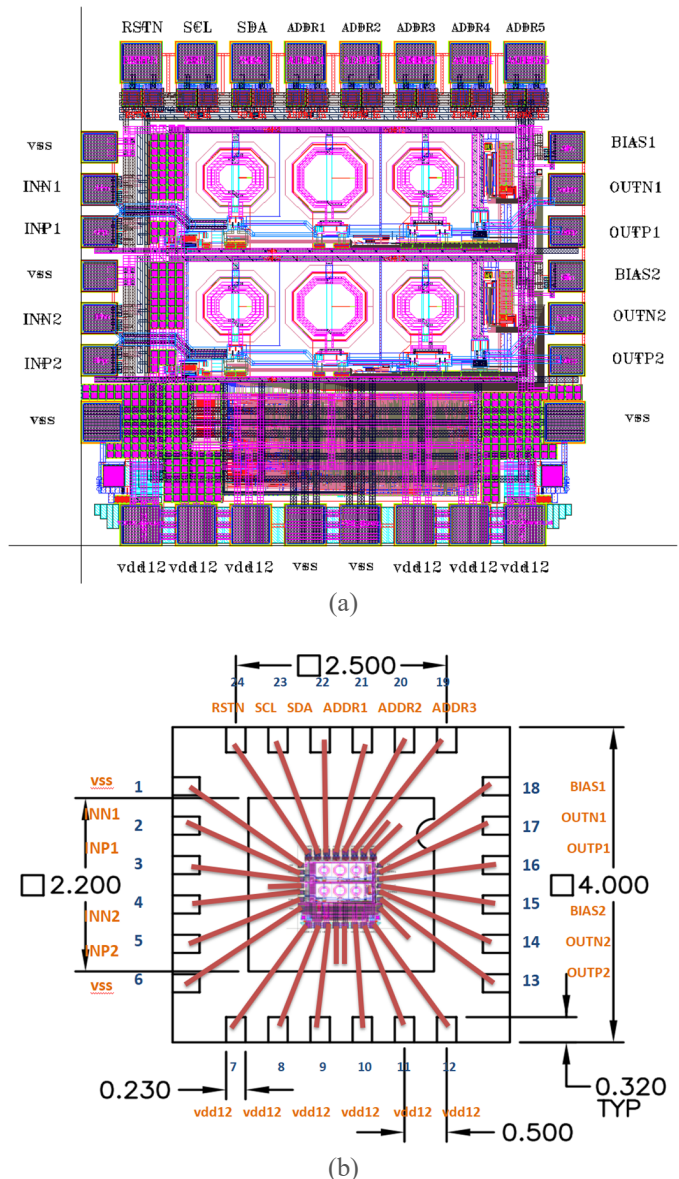

Fig. 5. Layout (a) and bonding diagram (b) of LOCld65.

### C. The output driver

The output driver is a fully differential amplifier with two 50-Ω pull-up resistors as the load and the termination, as shown in Fig. 4(a). This output stage also employs the CTLE technique with a 3-bit programmable resistor to tune the peaking strength. Fig. 4(b) shows the post-layout AC simulation results of the output modulation current. After the parasitic parameters of output pads and bond wires are considered, the bandwidth is from 11.3 GHz to 14.6 GHz at the typical case with different resistance settings. The modulation current for the VCSEL is adjustable by tuning the tail current of the output driver. The range of the modulation current amplitude is greater than 8 mA at the worst process corner.

### D. $I^2C$ slave

LOCld65 has 32 bits of internal registers to control the input equalization strength, the feedback strength, the bias current, the modulation current, and the output equalization strength of each channel. LOCld65 implements an $I^2C$ slave to configure the internal registers. When the $I^2C$ slave is reset, the internal registers are configured with default settings. With an external power-on-reset circuit, the ASIC operates at power up. In order to protect the internal registers and the $I^2C$ slave from the radiation-induced Single Event Upsets (SEUs), the $I^2C$ slave is implemented with triple modular redundancy.

### E. Package

LOCld65 has an area of 1 mm x 1 mm. Fig. 5(a) is the layout of the die. LOCld65 is packaged in a 24-pin open-cavity Quad-Flat No-leads (QFN) package. Fig. 5(b) is the bonding diagram. The packaged chip is 4 mm x 4 mm. The die is located in the center of the package. There are 24 pins in total and the pitch of the pins is 0.5 mm.

## III. TEST RESULTS OF LOCLD65

Four prototype chips have been tested for eye diagrams and one chip has been tested in x-rays for Total Ionizing Dose (TID) test. An SEU test will be conducted in the future.

### A. Test setup

In order to test LOCld65, we improved the LC-TOSA-based optical transmitter module MTx that was designed for the ATLAS Liquid (LAr) Argon Calorimeter Phase-I trigger upgrade. The improved optical transmitter is renamed as MTx+ [26]. Comparing to MTx, we made the two improvements. First, we used the standard Enhanced Small Form-factor Pluggable (SFP+) connector as the electrical interface and the standard Lucent Connector (LC) as the optical interface. Second, we used a custom optical latch to hold two LC connectors and the two TOSAs together. We used a 3D printer to print optical latches for the moment and will produce on injection molding in the future. The module dimension is 44.5 mm (length) x 18.2 mm (width) x 5.8 mm (height). The module stays below 6 mm to meet the mechanical requirement in the ATLAS LAr calorimeter and can be both panel and board mounted. We use two TOSAs from Truelight (Part No. TTF-1F59-427). Fig. 6(a) and Fig. 6(b) show the 3D bottom view of an MTx+ module with two TOSAs inserted in a latch and assembled to a PCB and the 3D top view of an MTx+ module plugged into a motherboard. Fig. 6(c) is a picture of an MTx+ module plugged into a test motherboard.

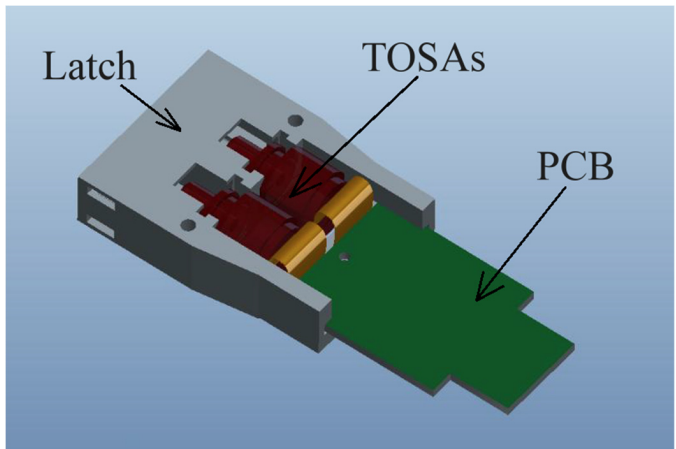

(a)

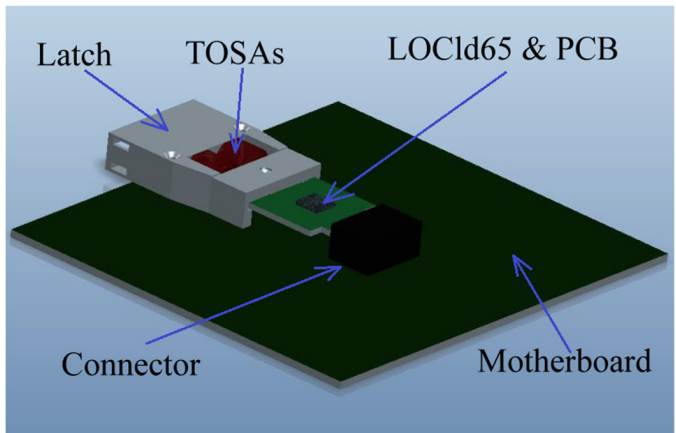

(b)

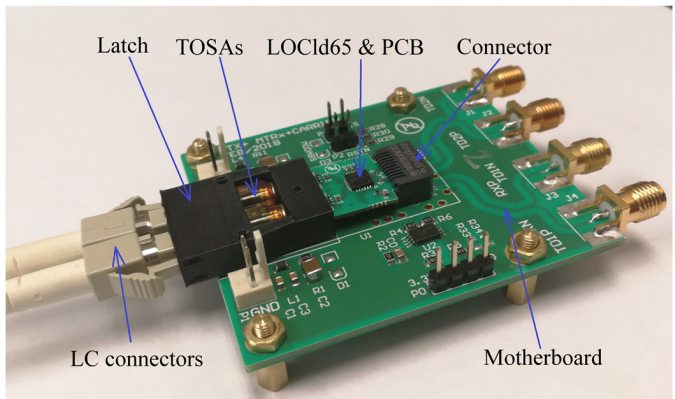

(c)

Fig. 6. Bottom view (a) and top view (b) of an MTx+ module with an optical latch and 2 TOSAs and a picture of an MTx+ module plugged in the test motherboard (c).

Fig. 7(a) is the block diagram of the test setup. A pattern generator (Model No. SDG 12070 produced by Picosecond Pulse Lab) provided a pair of pseudorandom binary sequence ($2^7-1$) differential signals. A pair of 10-dB attenuators attenuated the input signals to the minimum input amplitude of 100 mV (P-P). The output of MTx+ fed a sampling oscilloscope (Model No. TDS 8000B with an optical module 80C08C produced by Tektronix) through a 2-meter multiple-mode fiber to observe eye diagrams. The pattern generator provided a trigger signal for the oscilloscope. A power supply provided 1.2 V and 3.3 V for MTx+. A picture of the test setup is shown in Fig. 7(b).

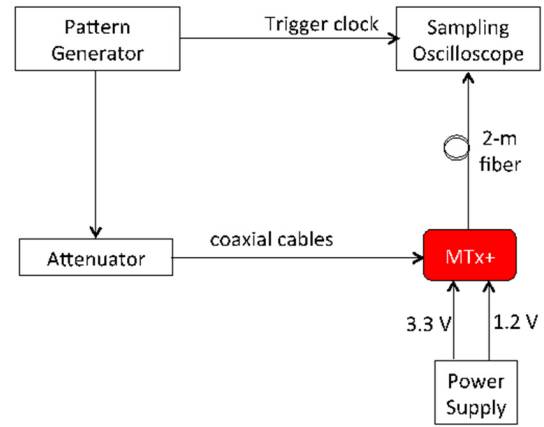

(a)

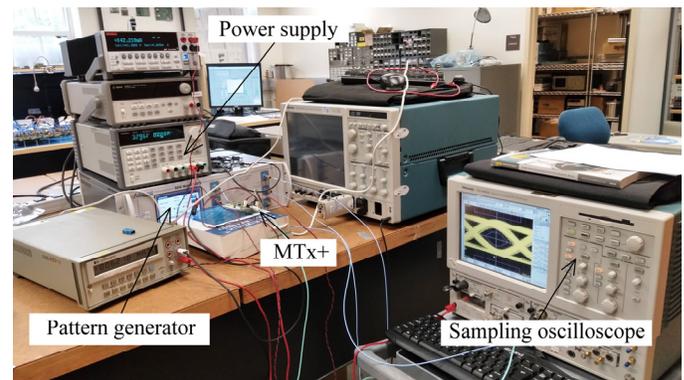

(b)

Fig. 7. Block diagram (a) and picture of the test setup (b).

B. Eye diagram test

Four prototype modules were tested for eye diagrams. The coaxial cables were 10 centimeters long. At 10 and 14 Gbps, all channels of all modules under test passed the eye mask test. Sample eye diagrams at 10 Gbps and at 14 Gbps are shown in Fig. 8(a) and Fig. 8(b), respectively. Shown in the figure are the eye masks of 10-Gbps Fiber Channel standard built in the oscilloscope. At 14 Gbps, the horizontal scale of the oscilloscope was adjusted to match the eye mask. The power currents of 1.2 V and 3.3 V of each channel are 38.8 mA and 6.6 mA, respectively. The power consumption at 14 Gbps is 68.3 mW/channel (the VCSEL included) with the VCSEL voltage of 3.3 V. Rise time and fall time are 35 ps and 45 ps, respectively at 14 Gbps.

We tested crosstalk of LOCld65. During the test, both channels operated at 14 Gbps, one channel serving as the aggressor and the other as the victim. When the aggressor channel was turned on or off, no significant effect was observed in the victim channel.

## C. Input equalizer test

In order to demonstrate the function of the input equalizer, we changed the coaxial cables from 10 centimeters to 1 meter. Fig. 9(a) and Fig. 9(b) are the eye diagrams at 14 Gbps when the equalization strength was set to be minimum and moderate, respectively. The diagrams demonstrate that by adjusting the input equalization strength, the high-frequency loss of long cables can be compensated.

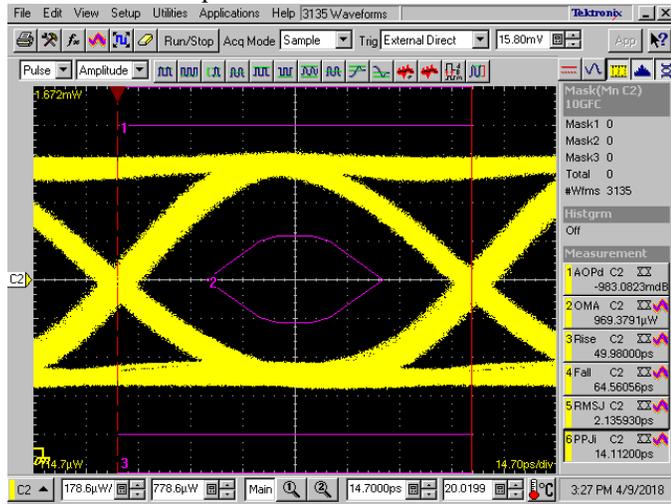

(a)

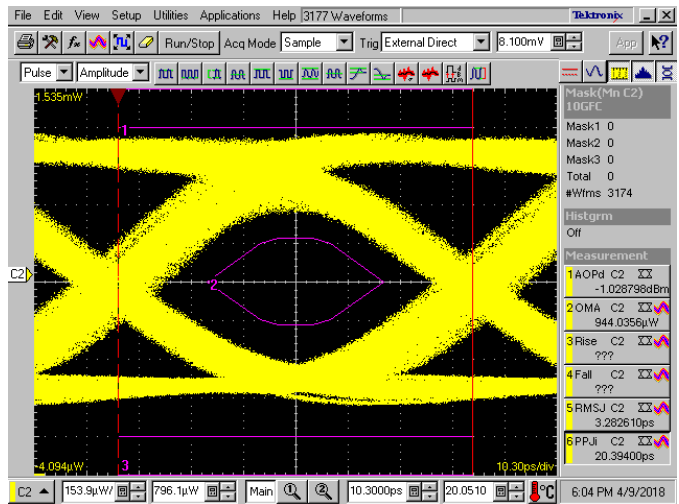

(b)

Fig. 8. Eye diagrams at 10 Gbps (a) and at 14 Gbps (b).

## D. Headroom test

A common concern in high-energy physics applications, where the power voltage decreases over time, is the voltage headroom of a VCSEL because its forward voltage may depend on temperature and radiation damage [27]. We tested the voltage headroom of LOCld65. Fig. 10 is an eye diagram at 14 Gbps with a VCSEL voltage of 2.5 V. The power currents of 1.2 V and 2.5 V are 38.8 mA and 6.2 mA, respectively. With the VCSEL voltage of 2.5V, the power consumption is 62.1 mW/channel. The test results show that at 14 Gbps, the VCSEL voltage can be as low as 2.1 V.

## E. Irradiation test results

A prototype chip was tested in x-rays. We used the same setup as shown in Fig. 7(a). Fig. 11 is a picture of the test setup. During the test, the MTx+ module was located in an x-ray irradiator (Model No. XR-160 produced by Precision X-Ray, Inc.), whereas all the other components were protected. The eye diagrams and the power currents were monitored during the test. The maximum x-ray energy was 160 keV. The dose rate was 1.3 Gy($SiO_2$)/s. The tested lasted 65 minutes and the TID reached 4.9 kGy($SiO_2$).

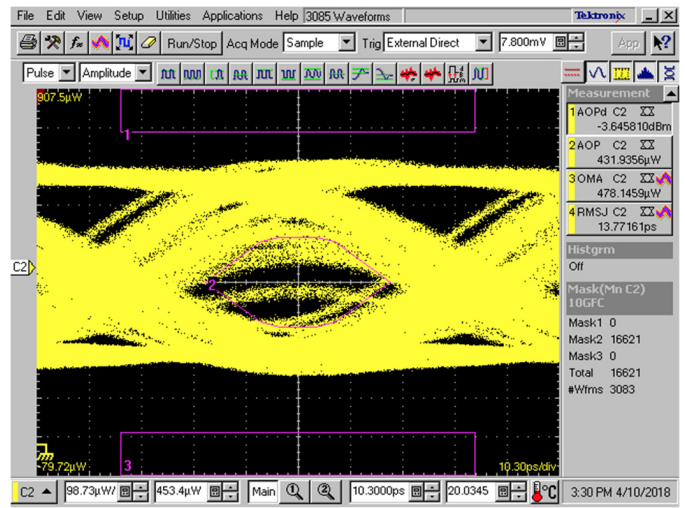

(a)

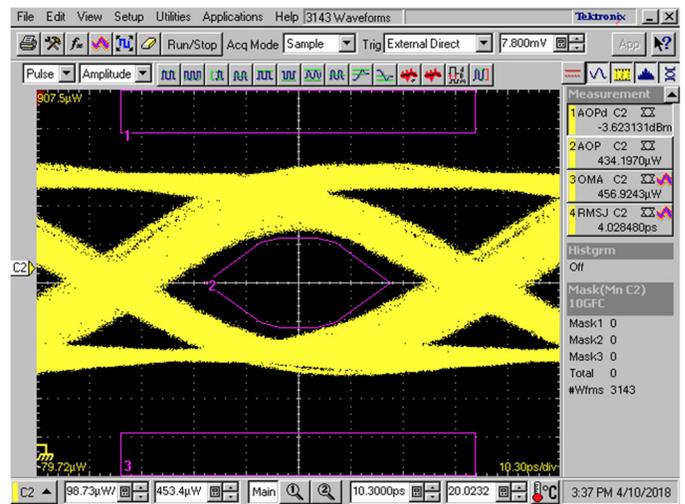

(b)

Fig. 9. Eye diagram without input equalization (a) and with input equalization (b) at 14 Gbps.

Fig. 12(a) and Fig. 12(b) show the relative power current change and the performance change of LOCld65 during irradiation test. The core current (1.2 V) and the bias current (3.3 V) decreased by 1.0% and 2.3%, respectively. The Optical Modulation Amplitude (OMA) and the Average Optical Power (AOP) decreased by 2.4% and 3.6%, respectively. No significant change was observed in rise time, fall time, and jitter (RMS).

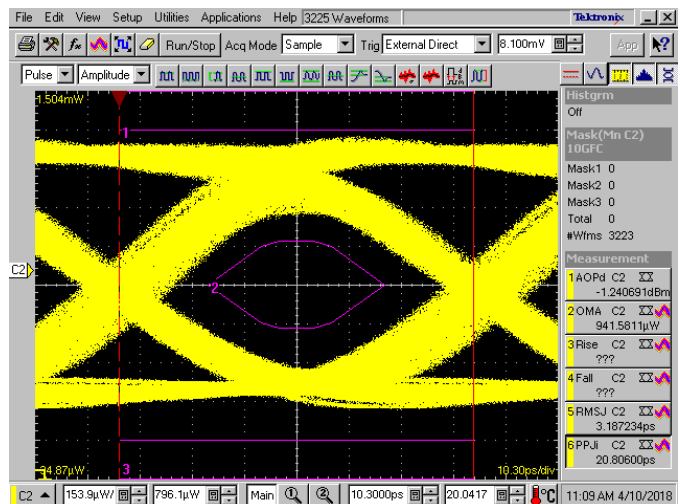

Fig. 10. Eye diagram at 14 Gbps with a 2.5-V VCSEL voltage.

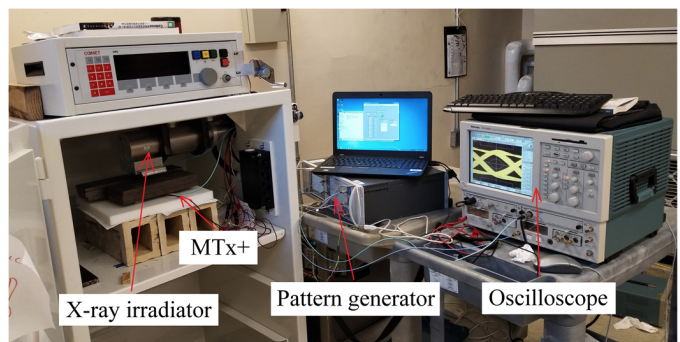

Fig. 11. Picture of the irradiation test setup.

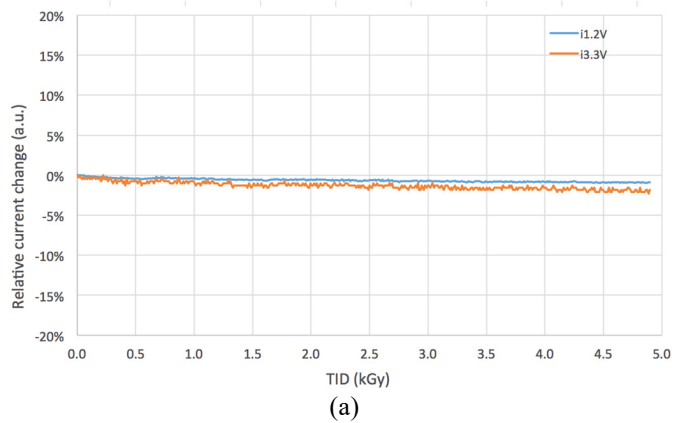

(a)

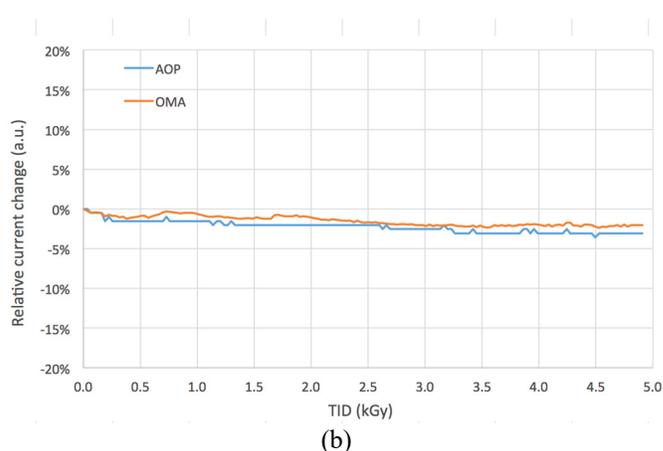

(b)

Fig. 12. Relative current change (a) and relative performance change (b) during the irradiation test.

## IV. Conclusion

We present the design and the test results of a dual-channel Vertical-Cavity Surface-Emitting Laser (VCSEL) driver ASIC LOCld65 for detector front-end readout. LOCld65 is designed in a commercial 65-nm CMOS technology with a power supply of 1.2 V. LOCld65 contains two separate channels with the same structure and the two channels share an I$^2$C slave. Each channel consists of an input amplifier, four stages of limiting amplifiers (LAs), a high-current output driver, and a bias-current generator. In order to extend the bandwidth, the input amplifier uses an inductive peaking technique and the LAs use a shared inductive peaking technique. The input amplifier and the output driver each utilize a CTLE. The LAs employ active feedback. The modulation current, the bias current, the peaking strength of the CTLEs, and the feedback strength of LAs are programmable through an I$^2$C interface. In order to protect from the radiation damage, the I$^2$C slave is implemented with triple modular redundancy. LOCld65 is tested to operate up to 14 Gbps for each channel with typical power dissipations (the VCSEL included) of 68.3 mW/channel and 62.1 mW/channel at the VCSEL voltages of 3.3 V and 2.5 V, respectively. LOCld65 survives 4.9 kGy(SiO$_2$). LOCld65 is an excellent match for lpGBT in single- or dual-channel optical transmitters in HL-LHC upgrade applications.

## Acknowledgment

We are grateful to Drs. Paulo Moreira, Syzmon Kulis, and Sandro Bonacini from CERN for their help in the I$^2$C slave design.